\documentclass[aps,prb,twocolumn,showpacs,superscriptaddress,floatfix]{revtex4-1}

\usepackage{amsmath}
\usepackage{amssymb}
\usepackage{graphicx, hyperref}
\usepackage{grffile} 
\usepackage{float}
\usepackage{color}
\usepackage{subfigure}
\usepackage{soul}
\hypersetup{draft}
\usepackage{lipsum}

\newcommand{\tb}[1]{\textbf{#1}}

\newcommand{\kk}{\mathbf{k}}
\newcommand{\qq}{\mathbf{q}}

\newcommand{\unitW}{$\mathrm{eV}$ }

\begin{document}

\author{A.~Kumar}
\email{akumar13@ncsu.edu}
\affiliation{Department of Physics, North Carolina State University, Raleigh, North Carolina 27695, USA}

\author{A.~F.~Kemper}
\email{akemper@ncsu.edu}
\affiliation{Department of Physics, North Carolina State University, Raleigh, North Carolina 27695, USA}

\title{Higgs Oscillations in time-resolved Optical Conductivity}

\begin{abstract}
 Driving superconductors out of equilibrium is a promising avenue to study their equilibrium properties as well as to control the superconducting state. Non-equilibrium superconductors are often studied using time resolved optical conductivity measurements. Thus, the characterization of a superconducting state in a pump driven non-equilibrium state requires careful attention in the time domain. We calculate time-resolved optical conductivity of a pumped superconducting state using a non-equilibrium Keldysh approach. Through functional derivation, the optical conductivity is obtained with full vertex corrections and used to characterize the transient superconducting state. The transient optical conductivity shows the suppression of the superconducting order parameter in the time domain. The subsequent recovery of the order parameter exhibits oscillatory behavior that corresponds to the Higgs amplitude mode, and may be seen in several parts of the spectrum.

 \end{abstract}

\maketitle
 \section{Introduction}
 Advancements of time-resolved spectroscopic techniques have enabled the measurement \cite{2011Graf} and control
 \cite{2011Fausti, 2014Hu, 2016Mitrano, 2018aYang, 2018bYang, 2018cYang} of hidden properties of ground states and low
 energy excitations in correlated materials, which are not easily accessible in equilibrium. In the energy spectrum,
 relatively low energy scale -- terahertz and mid-infrared -- frequencies have a special place in such techniques
 because of their non-invasive nature (to suppress the heat production that may destroy the ordered state) and
 suitability for observation of low energy excitations of the ground states. This advantage enables the characterization
 of the ordered ground state and study of the relaxation of the low energy excitations.

 Within the context of superconductivity, the excitations of superconducting condensate are of great interest.
 Experimentally, it is difficult to excite a superconducting order because the order parameter is a scalar field
 (spinless and chargeless), and does not couple to a vector electromagnetic (EM) field in the linear regime
 \cite{2015Pekker}. Because a superconducting state exhibits broken $U(1)$ gauge symmetry, when the order is perturbed, $U(1)$
 symmetry breaks spontaneously and results in two oscillating bosonic modes: first, the massless phase mode through
 Goldstone-Nambu mechanism and second, the massive amplitude mode through Higgs-Anderson mechanism. The phase mode gets
 pushed to the plasma frequency range because of coupling to Coulomb interactions. The other excitation, the Higgs mode
 resides at a frequency of twice the superconducting gap ($2\Delta$). Recently, it has been shown that the amplitude
 mode can be excited, and observed, by an EM field using a coexisting order e.g. charge density wave \cite{2014Measson},
 by non-linear coupling to EM field (generation of third harmonic) \cite{2014Matsunaga}, by time-resolved conductivity
 \cite{2013Matsunaga} or by using the presence of a supercurrent \cite{2018Nakamura, 2017Moor}.  

 Given that the optical conductivity is the primary probe of the Higgs mode, a proper theoretical description of the
 conductivity is a necessity.  Numerous work has been done to calculate the conductivity (and other response functions)
 of correlated electrons in an equilibrium \cite{1958Mattis, 1991Zimmermann, 1993Chen, 2017Chubukov} and non-equilibrium
 state\cite{ 2008Eckstein, 2010Eckstein, 2015Orenstein, 2014Krull}. In a non-equilibrium state, such as induced by a
 pump-probe setup, a calculation of the conductivity through the Bethe-Salpeter (BS) equation, which is necessary to
 capture effects beyond the bare-bubble susceptibility, becomes computationally prohibitive because the Hamiltonian loses
 time-translational symmetry, resulting in 4 separate time variables in the BS equation.  Previous solutions to this
 problem includes exact diagonalization of the Hamiltonian (which is limited by system size) and mean-field analyses of
 the BCS Hamiltonian (which is a priori not gauge invariant, and neglects inelastic collisions and dynamics of the
 interactions) \cite{1960Nambu}. A recent proposal -- dotted tdDMFT\cite{2016Tsuji} --  has been used to calculate the
 pair field susceptibility of a superconductor. However, the tdDMFT method is only fully justified for an infinite
 dimensional case.

 In this work, we go beyond these limitations and calculate the time-dependent optical conductivity using a functional
 derivation approach based on non-equilibrium Green's functions.  We solve the Nambu-Gor'kov equations for
 electron-phonon mediated superconductivity self-consistently on the Keldysh contour and evaluate the non-equilibrium
 interacting Green's functions in the time domain. We consider the Holstein model with impurity scattering as a
 particular instance to study the transient optical conductivity of a superconductor. The optical conductivity is
 calculated by a functional derivative of the current with respect to the applied field. One of the advantages of this
 particular method is that it naturally includes vertex corrections \cite{2012aSakkinen, 2012bSakkinen,
 book_stefanucci}, but bypasses the calculation of the BS equation in the time domain. In equilibrium, our results
 reproduce several features of the known conductivity of dirty superconductors\cite{1958Mattis,1991Zimmermann} such as
 an upturn towards low frequencies inside the gap.  In a pump driven, non-equilibrium case, the conductivity reflects
 the temporal dynamics of superconducting order including suppression, recovery, and the Higgs oscillations. These are
 clearly present in the features of the conductivity that are commonly associated with the superconducting order in
 equilibrium, i.e. the energy location of the gap, the coherence peak, and the phonon features in the real part of the
 conductivity, as well as in the inductive $1/\omega$ low-frequency response in the imaginary part. We quantify and
 characterize the transient superconducting state using these features as well a purely time domain feature (the probe
 current). All the quantities show excellent correlation to the gap dynamics which are known from the underlying
 simulations\cite{2015Kemper}. 
 
 The structure of the paper is as follows: first, we describe the Hamiltonian and the method to calculate optical
 conductivity in equilibrium and non-equilibrium. Then, we discuss the results for the equilibrium state and the
 dynamics of the Cooper pairs in a pump-driven non-equilibrium state. 
 
\section{Methods}
 We use a minimal Hamiltonian to simulate a phonon-mediated, $s$-wave superconductor. We consider the Holstein
 Hamiltonian on 2D square lattice 
 \begin{align}
     \mathcal H &= \sum_{\kk,\sigma}\xi(\kk) c^\dagger_{\kk,\sigma} c^{\phantom\dagger}_{\kk, \sigma} + \Omega  \sum_{\qq}
                             \left(b_{\qq}^\dagger b^{\phantom\dagger}_{\qq} + \frac{1}{2} \right) \nonumber\\
               &+\frac{g}{\sqrt{N}}\sum_{\substack{\sigma\\\kk,\qq}}c_{\kk+\qq, \sigma}^\dagger
         c^{\phantom\dagger}_{\kk,\sigma} \left( b^{\phantom\dagger}_{\qq} + b_{-\qq}^\dagger \right)  \nonumber \\
         &+ \sum_{i} V_i c_i^\dagger c_i
 \end{align}
 Here, $\xi(\kk)$ ($= -2 \mathrm{V_{nn}} \left[ \cos(k_x) + \cos(k_y) \right] -\mu$) is the nearest neighbor tight-binding
 energy dispersion measured relative to the chemical potential, $c^\dagger_\kk, c^{\phantom\dagger}_\kk$ ($b^\dagger_\qq,
 b_\qq)$ are the standard creation and annihilation operators for an electron (phonon), $g$ is the momentum-independent
 $e$-ph coupling constant, and $\Omega$ is the frequency for the Einstein phonon. $V_i$ is the coupling between electrons
 and impurities which are distributed randomly on lattice sites. 

 The phonon subsystem is treated as a heat reservoir whose properties do not change in time as we drive the electronic
 subsystem, which is valid for the small pump fluences considered here \cite{2018Abdurazakov, 2015Murakami}.
 The interactions are treated within a self-consistent perturbative framework, which sums the diagrams to all orders.
 The self energy is calculated using from a Luttinger-Ward functional to ensure conservation-laws \cite{1961Baym}. We
 use the first order diagram for the electron-phonon interaction, and the self-consistent Born approximation for the
 impurity scattering. The superconducting state is treated within a self-consistent Migdal-Eliashberg formalism, and
 the time evolution is done by solving the Gor'kov equations self-consistently on the Keldysh contour \cite{2015Kemper}. 

 Table~\ref{tab:parameters} lists the parameters used in the simulation; these parameters result in an equilibrium
 superconducting gap $\Delta \approx 46$ meV. The choice of parameters does not represent a specific material. Rather,
 the parameters were chosen for numerical tractability. Real systems can be simulated by appropriately rescaled
 parameters. 
 
 The pump field, which is applied in the $(11)$ direction, is included to all orders via Peierls' substitution. In
 addition to the pump pulse, a secondary (probe) pulse is included in the same way as the pump. However, the probe
 amplitude and frequency are optimized to ensure that the probe is in the linear-response regime and is able to probe the
 conductivity within the $2\Delta$ range. The pump and probe pulse envelopes are taken to be Gaussian curves
 $\mathrm{|\vec{A}(t)| = A_{max}cos(\omega t)\exp(\frac{-t^2}{2\sigma})}$.

\begin{table}[t]
    \caption{Parameters used in the simulation}
    \label{tab:parameters}
    \begin{ruledtabular}
        \begin{tabular}{lcr}
            Phonon frequency ($\Omega$) & $ 0.2$ \unitW \\
            Phonon coupling ($g^2$) &$0.12$ \unitW \\
            Impurity coupling ($|V_i|$) & 0.01 \unitW \\
            \hline
            Band parameters &\multicolumn{1}{c}{ $V_{nn}=0.25 \, \mathrm{eV}$, $\mu = 0.0 \, \mathrm{eV}$} \\
            Temperature &\multicolumn{1}{c}{ $\approx 83 \, \mathrm{K}$ } \\
            Pump pulse &\multicolumn{1}{c} { $\omega_p = 1.5\,\mathrm{eV}$ , $\sigma_p = 8 \; \mathrm{[1/eV]}$ } \\
            Probe pulse &\multicolumn{1}{c} { $\omega = 0.1\, ,  \sigma = 2 \; \mathrm{[1/eV]}$ } \\
    \end{tabular}
    \end{ruledtabular}
\end{table}

To calculate transient conductivity we have used the algorithm proposed in Refs. \cite{book_stefanucci, 2016Shao}.
The central idea is that first we calculate non-equilibrium current $\mathrm{\vec{J}_{pump}}$ for the pumped state
without a probe as follows:
$$\mathrm{ \mathbf{j}(t) = N_\kk^{-1} \sum_\kk \mathbf\nabla \xi(\kk-\vec{A})\ \mathrm{Im}\ G_\kk^<(t,t) } $$
where the derivative is taken along the field $(11)$ direction.  Then, for each pump-probe delay time ($\mathrm{t}$) we
calculate change in the current as function of time ($\delta \vec{J} =  \mathrm{\vec{J}_{pump+probe} -
\vec{J}_{pump}}$). The current and the probe time profiles are then used to calculate time-dependent conductivity as
$\mathrm{ \sigma(t,\omega) = \frac{\delta \vec{J}(t, \omega)}{\vec{E}_{probe}(\omega)} }$.  Here, we have taken the Fourier transform
along average-time axis ($t=t'$). However, depending on the experimental settings other time axes can also be used to
take the Fourier transform, as described in Refs. \cite{1995Hamm, 2002Nemec, 2002Averitt, 2005Nienhuys}. For this
particular choice, length of the time signal average outs the amplitude of the Higgs mode. We limits the time signal
length to $0.01$ [1/eV] to capture the Higgs oscillations within the numerical accuracy.

\section{Results}
    \subsection{Equilibrium}
\begin{figure}[]
            \includegraphics[width=0.48\textwidth, trim=0.2cm 0.2cm 0.2cm 0.2cm, clip=true]{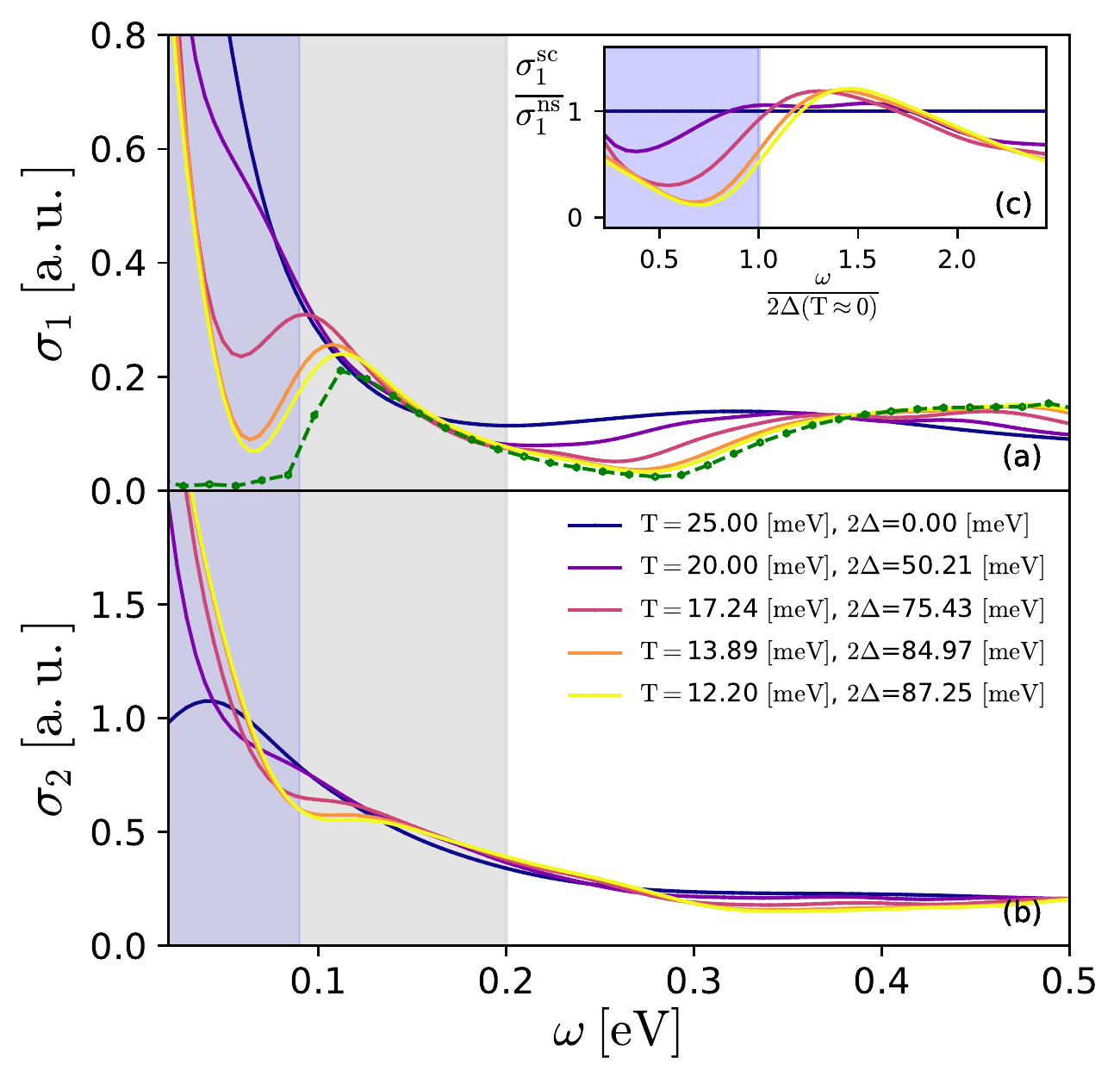}
            \caption{
                (Color online) The conductivity of the Holstein model in equilibrium. Panels \textbf{(a), (b)} show the
                real and the imaginary part of optical conductivity at different temperatures, respectively. The
                temperature-range spans the superconducting phase transition. The dark shaded region shows the maximum
                superconducting optical gap near zero temperature $2\Delta(T \approx 0)$. The light shaded region shows
                the phonon window ($0 < \omega <\Omega = 0.2$ eV).  Panel \textbf{(c)} displays the ratio of real parts
                of conductivity in the superconducting state to normal state at $T=25$ meV. The dashed line in panel
                \tb{(a)} shows $\sigma_1$ calculated using \textit{bare-bubble} susceptibility at $\mathrm{T} = 12.2$
                [meV].
                }
            \label{fig:equilibrium01}
    \end{figure}

    First, we calculate the conductivity of the system in equilibrium state, i.e. without a pump field. The results are
    shown in Fig.~\ref{fig:equilibrium01} as a function of temperature $T$. For reference, we have also labeled the
    curves by their equilibrium superconducting gap $2 \Delta$ as determined from the static component of the anomalous retarded
    self-energy $\Sigma^F_R(\omega=0)$. In the normal state ($T>T_c$) we observe the Drude features in the conductivity near
    zero frequency, and the effect of the Einstein phonon at the phonon frequency $\Omega$. The presence of a phonon
    lowers the optical spectral weight in the vicinity of the phonon frequency ($\Omega$); this may be observed as a
    flattening of the spectral weight in $\sigma_1$ at $\Omega$. It is important to note that the minimum of the real part
    of the conductivity lie at the phonon frequency $\Omega$ in the normal state and shifts by $2\Delta$ in the
    superconducting state. This particular feature will be used to study the dynamics of the superconducting edge (the
    gap) in the superconducting state.
    
    In the superconducting state ($T<T_c$) we observe the opening of the gap in the conductivity
    i.e. the lowering of the optical spectral weight inside the $2\Delta$ window (marked by the first shaded region in the
    figure near zero frequency). In addition, we note the shift of the minimum around the phonon frequency from $\Omega$
    to $\Omega + 2\Delta$. The normalized conductivity ($\sigma_1(T<T_c)/\sigma_1(T=25)$) is plotted in
    inset \tb{(c)}, which clearly shows the opening of the gap and the development of the coherence peak.
    The imaginary part of the conductivity (Fig.~\ref{fig:equilibrium01}\tb{(b)}) can also be analyzed in a similar way to
    characterize the superconducting state; it shows $\frac{1}{\omega}$ behavior inside the $2\Delta$ window. For comparison we include the conductivity calculated from the bare-bubble susceptibility
    $\chi(\qq=0,\omega) = \int d(t'-t'') \sum_\kk |\vec{v}_\kk|^2 [G^>_\kk(t',t'')G^<_\kk(t'',t') - G^<_\kk(t',t'')
    G^>_\kk(t'',t')]e^{-i\omega (t'-t'')}$ in the figure (dashed line). We observe noticeable qualitative differences
    because of the vertex corrections in the conductivity calculated from the functional derivation of probe current,
    mainly near the gap edge and for low energies. 

\begin{figure}[]
            \includegraphics[width=0.48\textwidth, trim=0.05cm 0.05cm 0.01cm 0.2cm, clip=true]{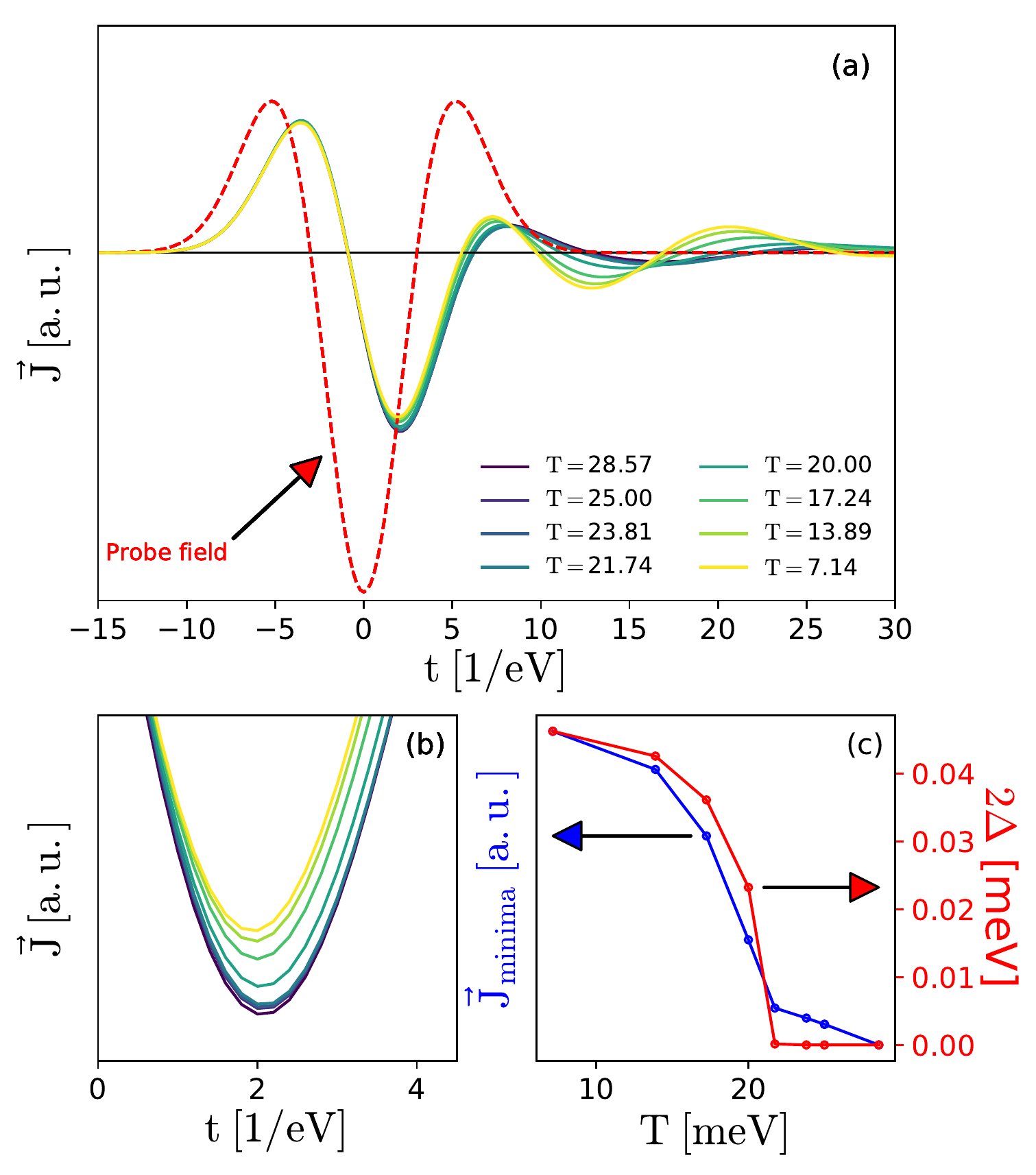}
            \caption{
                (Color online) The superconducting phase transition indicators. Panel \tb{(a)} shows probe current
                as a function of time at different temperatures in equilibrium. The temperature-range spans the
                superconducting phase transition. Panel \tb{(b)} displays the zoomed region around the global minimum of the
                probe current. Panel \tb{(c)} shows the correlation between the probe-current minimum and the superconducting
                order parameter calculated using $\mathrm{\Sigma^F_R(\omega=0)}$.
                }
            \label{fig:equilibrium02}
    \end{figure}

    As has been shown in THz pump-probe experiments \cite{2013Matsunaga}, the probe current maximum/minimum can
    also be used as an indicator to study changes in the physical state of a system, e.g. phase transitions or pump induced
    changes in the system response. We analyze our data in a similar way and plot the equilibrium probe current at different
    temperatures ranging from normal state to the superconducting state in Fig.~\ref{fig:equilibrium02}.
    As the temperature is reduced, rapid oscillations appear in the probe current, and the minimum is reduced (see Fig.~\ref{fig:equilibrium02}\tb{(b)}).
    The minimum of the probe current directly correlates with the superconducting order parameter, which is shown in 
    panel \tb{(c)}  where we plot the first minimum of the probe current and
    the superconducting order parameter. This particular feature can also be used to characterize the
    transient conductivity in a pump driven superconductor which is shown in the following sections.

    \subsection{Non-equilibrium}
\begin{figure}[]
        \includegraphics[width=0.48\textwidth, trim=0.2cm 0.2cm 0.2cm 0.2cm, clip=true]{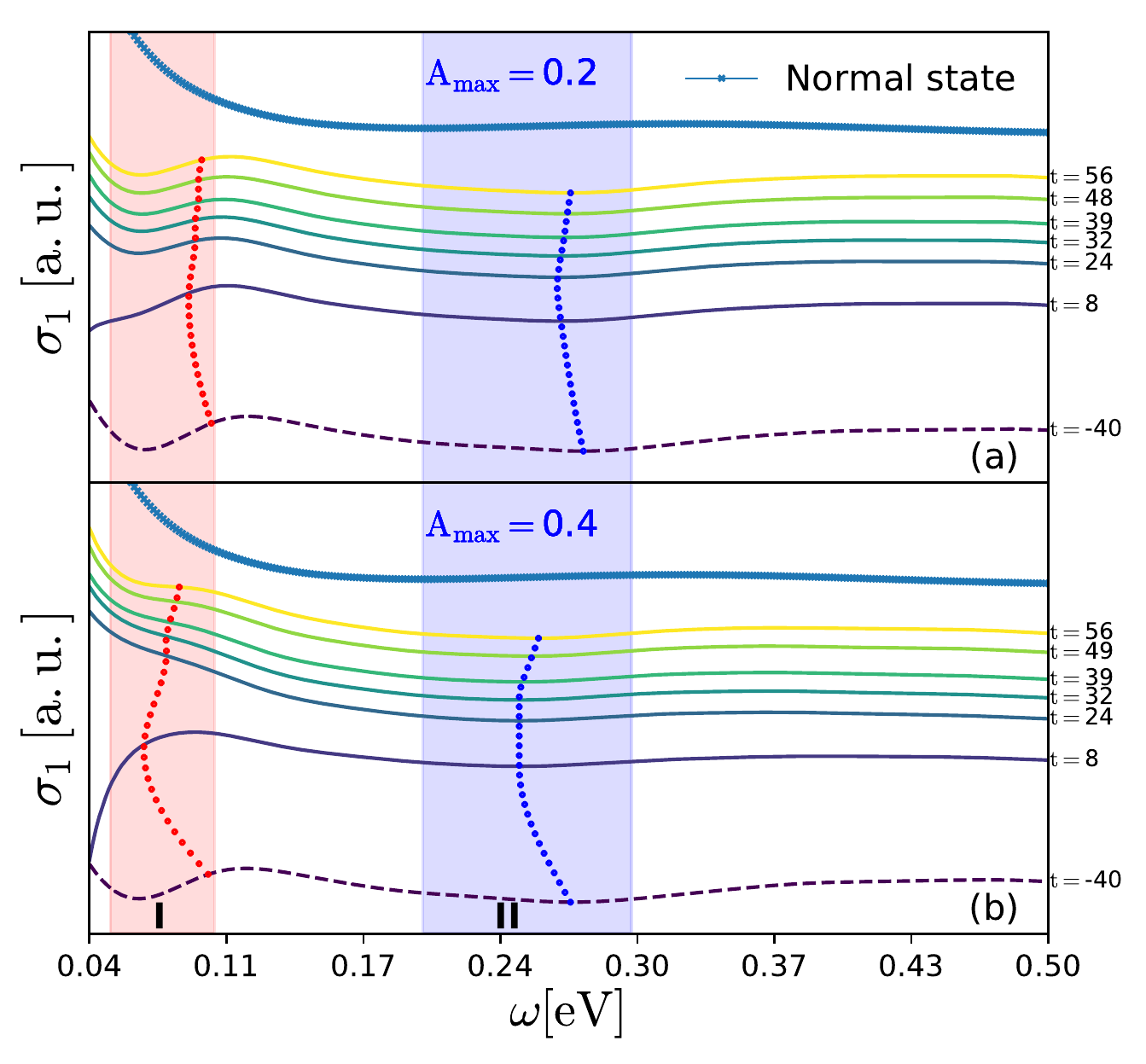}
        \caption{
            (Color online) Conductivity of the system in a pump driven non-equilibrium state. Panel \tb{(a)} and \tb{(b)}
            show the real part of optical conductivity as a function of frequency at different delay times for pump
            fluence $\mathrm{A_{max}=0.2 \; [1/a_0]}$  and $\mathrm{0.4 \; [1/a_0]}$. Each curve is shifted by an 
            offset (scaled as the delay time) along Y-axis to show the changes in the transient conductivity. The
            top dashed line in each panel shows the conductivity of normal state in equilibrium. Conductivity of the superconducting
            state is shown by $\mathrm{t = -40 \; [1/eV]}$ in equilibrium. The blue shaded region \tb{II} shows the
            vicinity where the location of the $\sigma_1$ minimum resides near the phonon frequency $\Omega$. The minima
            are shown by the dotted curve connecting different delay-time curves. The red shaded region \tb{I} shows the
            region where the gap edge is located and marked by the dotted line connecting different delay-time curves.
            }
        \label{fig:non-equilibrium01}
    \end{figure}

    In this section, we discuss the dynamics of electrons in a non-equilibrium state. We calculate the conductivity of a
    pump-driven non-equilibrium state. Figure~\ref{fig:non-equilibrium01} shows the real part ($\sigma_1$) of the
    transient conductivity as a function of pump-probe delay time for two pump fluences $\mathrm{A_{max}=0.2, 0.4 \;
    [1/a_0]}$. We observe noticeable changes in the conductivity from the equilibrium state (at $\mathrm{t=-40 \;
    [1/eV]}$). The suppression of superconducting order can be observed as the edge of the gap (indicated by red markers in region \tb{I})  moving towards zero
    during early delay times. As noted above, the location of the minimum in the real part of the conductivity near the
    phonon frequency $\Omega$ can also be used as an indicator of superconductivity. We observe that the minimum location also
    shifts on the frequency axis. These minima are located within the shaded region \tb{II} in the figure. Such
    suppression of superconductivity is expected in the transient state of the system when the pump drives the system
    because, intuitively, the pump injects energy in the system and creates excitations. These excitations raise the
    effective temperature of the system and result in the observed superconducting order suppression. It is important to
    notice that the effective-temperature picture does not imply a local equilibrium in the transient state as shown
    previously \cite{2015Kemper, 2017Nosarzewski, 2017Kumar}. Rather, the system is in a dynamic non-thermal state where
    oscillation of the superfluid condensate is observed (this will be discussed in the following sections). For the
    larger fluence $\mathrm{A_{max}=0.4 \; [1/a_0]}$, the melting of the superconducting order is stronger and the
    gap is closed further. Furthermore, it is important to notice that for larger fluences the conductivity ($\sigma_1$)
    does not show a "clear" spectral gap in the spectrum as shown in Figure~\ref{fig:non-equilibrium01} panel \tb{(b)}
    at delay time $\mathrm{t = 24[1/eV]}$. However, the superconductivity remains in the system (as evidenced by a
    finite off-diagonal order) as shown here \cite{2015Kemper}. Out of equilibrium, the direct connection between a gap
    in the optical spectra and finite off-diagonal order is broken.

    \subsection{Gap suppression and the Higgs oscillation}
    It has been proposed and observed \cite{2013Matsunaga, 2014Matsunaga} that when a superconducting condensate is
    perturbed by an ultrafast pump field, the relaxation of the excited population of the Cooper pairs follows an
    oscillatory decay mirroring the dynamics predicted for the order parameter\cite{1974Volkov, 1982Littlewood,
    2014Krull, 2015Cea, 2015Kemper, 2017Kumar, 2016Murakami}.  Such amplitude oscillations are called Higgs (or
    Anderson-Higgs) oscillations. Here they arise from a time dependence of the underlying order parameter, which is
    reflected in the time-dependent conductivity.

\begin{figure}[]
            \includegraphics[width=0.48\textwidth, trim=0.2cm 0.2cm 0.2cm 0.2cm, clip=true]{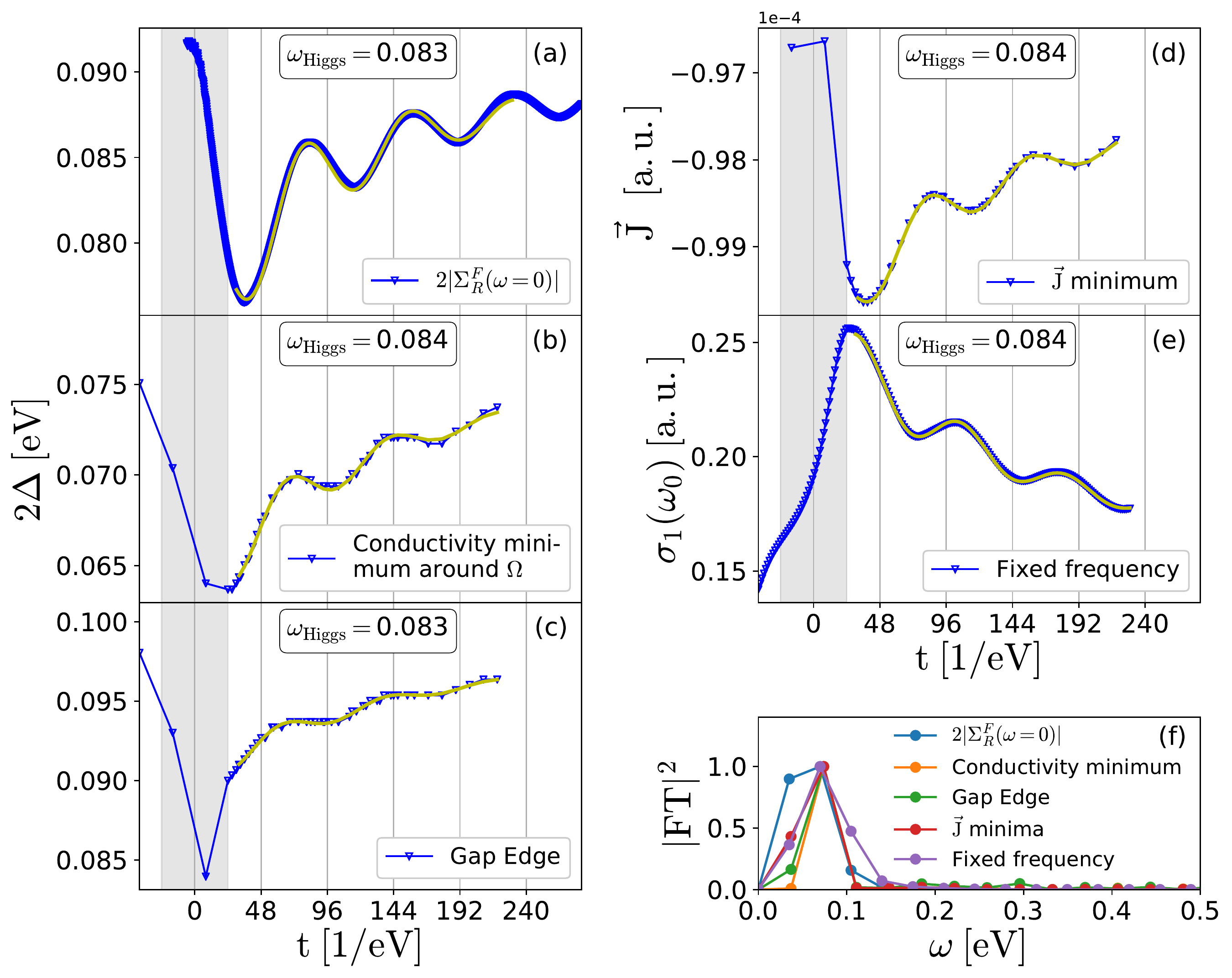}
            \caption{
                (Color online) \tb{The Higgs oscillation}. The figure shows the dynamics of the superconducting order
                parameter in a pump induced non-equilibrium state (for $\mathrm{A_{max}=0.2 \; [1/a_0]}$). Panel \tb{(a)}
                shows the order parameter as a function of average time calculated using anomalous self energy. Panels
                \tb{(b)} and \tb{(c)} display the estimated order parameter using the location of the minimum of the real
                part of optical conductivity around the phonon frequency (shown in the blue shaded region \tb{II} in
                figure~\ref{fig:non-equilibrium01}) and the gap edge location (the red shaded region \tb{I} in
                figure~\ref{fig:non-equilibrium01}), respectively. Panel \tb{(d)} and \tb{(e)} show the probe current
                minimum and $\sigma_1(\omega=0.09)$ as a function of time, respectively. All the five quantities shown in
                panels exhibit oscillations in time with approximately same frequency
                ($\omega_{\mathrm{Higgs}} \approx 2\Delta(t)$), which is calculated by fitting the data to an oscillatory
                decaying function. The Fourier transform of the data also shows a peak at the same frequency (panel
                \tb{(f)}).
                }
            \label{fig:suppression01}
    \end{figure}

    Here we study the dynamics of the superconducting order parameter in the pump-induced non-equilibrium
    state from the perspective of the time-resolved optical conductivity. For reference, we will use the superconducting
    gap $\Delta = \Sigma^F_R(\omega=0)$ as a function of average time. We may estimate the order parameter using the gap
    edge in $\sigma_1(\omega)$, which in equilibrium occurs at $\omega=2\Delta$. We define the edge of the gap as the
    point $\omega_{edge}$ on the frequency axis where the mean of $(\frac{\sigma_1^{sc}}{\sigma_1^{ns}})_{max} $ and
    $(\frac{\sigma_1^{sc}}{\sigma_1^{ns}})_{min}$ is located within the shaded region \tb{I} in the figure~\ref{fig:non-equilibrium01}. Similarly,
    we use the $\sigma_1(\omega)$ minimum around the phonon frequency $\Omega$ as a reference. These markers are shown in
    Fig.~\ref{fig:suppression01} panels \tb{(a-c)}, respectively.  We observe that the order parameter determined from
    $\Sigma^R_F$ is suppressed when the pump is active at early times, and it recovers back to the equilibrium value for
    later times, exhibiting Higgs oscillations as it recovers.  The gap edge and the minimum location exhibit similar
    behavior, although the relative change is larger at the minimum. All three quantities show Higgs oscillations at
    approximately the same frequency\textemdash in principle the frequency is time dependent as it scales with the local
    (in time) gap\cite{2015Kemper}, however in this time range is it approximately constant. Note that there is a small discrepancy in the
    order parameter value calculated using various pieces of the conductivity data which may arise due to the particular
    choice of delay axis to Fourier transform, and due to the frequency resolution of the probe
    signal.  
    
    Besides these markers, the Higgs oscillations occur across the response. For example, the probe-current minimum as a
    function of delay time show the Higgs oscillations as well (c.f. Fig.~\ref{fig:suppression01}d). Finally, the
    oscillatory behavior can also be observed when we considering $\sigma_1$ at some fixed frequency $\omega_0$ as shown
    in Fig.~\ref{fig:suppression01}e. Here we have chosen $\omega_0$ within $2\Delta$ window, but the oscillation of the
    conductivity may be seen at all frequencies as a function of time delay (panel \tb{(c)} Figure~\ref{fig:suppression01}).
    A movie of the time-resolved conductivity is available as a supplement.
   
   Panel \tb{(f)} presents the Fourier transform of the quantities shown in panels \tb{(a-e)}.  Although the limited data
   length leads to wide peaks, the various measurements all oscillate at the same frequency. This is further confirmed
   by a curve fit to a decaying oscillation (shown on the individual panels), which yields the same frequency for all the measures.
   
\section{Summary}
 We have presented the time-resolved optical conductivity for a pumped superconductor based on gauge invariant, fully
 vertex corrected method. The results show that the entire spectrum undergoes changes that reflect the underlying
 changes in the gap. There are shifts (in energy) of features in the conductivity due to the reduction from
 $\Delta_\mathrm{equilibrium}$ to some reduced $\Delta(t)$, which itself oscillates in time.  These ``Higgs''
 oscillations are thus visible in essentially the entire spectrum. We quantify several features that are known to
 correspond to the gap in equilibrium, e.g. the gap edge and the phonon minimum, and connect them to the underlying gap
 dynamics which are known from the calculations.

 In this work, our analysis is mainly based on the real part of the transient conductivity. For the Higgs mode matter,
 the imaginary part also exhibits similar oscillations. However, the imaginary part plays an important role in experimental
 data analysis. Recently \cite{2016Mitrano}, the equilibrium-like properties of the imaginary part were utilized to
 support the claim of the enhanced superconductivity observed in organic superconductor $K_3C_{60}$. It is important
 to understand whether transient optical conductivity exhibits such equilibrium-like properties. This should be analyzed in
 the future work. 
 
 Finally, we stress on the suitability of the method used in this work to calculate transient optical conductivity. The
 method enables calculation of the response functions beyond the bare-bubble susceptibility.
 The effect of vertex corrections varies depending on the particulars of the system.  For example, they are expected to
 be minor for an electron-phonon system in the Migdal limit, but not negligible when it comes to impurity scattering in
 certain regimes (this effect is observed in Fig.~\ref{fig:equilibrium01} for low energies where the impurity scattering
 is significant). The functional derivative method captures these faithfully and may have broader applicability in the
 evaluation of equilibrium and non-equilibrium two-particle quantities.
 
\section{ACKNOWLEDGMENTS}
We acknowledge fruitful discussion with Avinash Rustagi. This work was partly supported by NSF
DMR-1752713. This research used resources of the National Energy Research Scientific Computing Center, a DOE Office of
Science User Facility supported by the Office of Science of the U.S. Department of Energy under Contract No.
DE-AC02-05CH11231.

%

\end{document}